\documentclass[11pt]{article}

\usepackage{fullpage}
\usepackage{amsmath}
\usepackage{amsfonts}
\usepackage{amssymb}
\usepackage{amsthm}
\usepackage{times}

\newcommand{\ket}[1]{|\hspace{0.5pt}#1\hspace{0.5pt}\rangle}

\newcommand{\bra}[1]{\langle\hspace{0.5pt}#1\hspace{0.5pt}|}

\newcommand{\op}[1]{\operatorname{#1}}
\newenvironment{mylist}[1]{\begin{list}{}{
	\setlength{\leftmargin}{#1}
	\setlength{\rightmargin}{0mm}
	\setlength{\labelsep}{2mm}
	\setlength{\labelwidth}{8mm}
	\setlength{\itemsep}{0mm}}}
	{\end{list}}

\newtheorem{theorem}{Theorem}[section]
\newtheorem{lemma}[theorem]{Lemma}
\newtheorem{prop}[theorem]{Proposition}
\newtheorem{cor}[theorem]{Corollary}
\theoremstyle{definition}
\newtheorem{definition}[theorem]{Definition}

\newtheorem{remark}[theorem]{Remark}

\begin{document}

\title{\Large\bf Quantum Arthur-Merlin Games}

\author{
  Chris Marriott \quad\quad John Watrous\\[2mm]
  Department of Computer Science\\
  University of Calgary\\
  2500 University Drive~~NW\\
  Calgary, Alberta, Canada~T2N 1N4
}

\date{June 15, 2005}

\maketitle

\begin{abstract}
This paper studies quantum Arthur-Merlin games, which are Arthur-Merlin games
in which Arthur and Merlin can perform quantum computations and Merlin can send
Arthur quantum information.
As in the classical case, messages from Arthur to Merlin are restricted to
be strings of uniformly generated random bits.
It is proved that for one-message quantum Arthur-Merlin games, which
correspond to the complexity class $\mathrm{QMA}$, completeness and soundness
errors can be reduced exponentially without increasing the length of Merlin's
message.
Previous constructions for reducing error required a polynomial increase in 
the length of Merlin's message.
Applications of this fact include a proof that logarithmic length quantum
certificates yield no increase in power over $\mathrm{BQP}$ and a simple proof
that $\mathrm{QMA}\subseteq\mathrm{PP}$.
Other facts that are proved include the equivalence of three (or more) message
quantum Arthur-Merlin games with ordinary quantum interactive proof systems
and some basic properties concerning two-message quantum Arthur-Merlin games.
\end{abstract}


\section{Introduction}

Interactive proof systems and Arthur-Merlin games were introduced by
\cite{GoldwasserM+89} and \cite{Babai85} (see also \cite{BabaiM88}) in order
to model the notion of computationally efficient verification.
In an interactive proof system, a polynomial-time verifier with a private
source of uniformly generated random bits interacts with a computationally
unbounded prover in an attempt to check the validity of the claim that a
common input string is contained in some prespecified language.
Arthur-Merlin games are similar in principle to interactive proof systems,
but are somewhat more restricted---the verifier (called Arthur in this
setting) no longer has a private source of randomness, but instead has only
a public source of randomness that is visible to the prover (called Merlin).
Because Arthur is deterministic aside from the bits produced by the random
source, one may without loss of generality view that an Arthur-Merlin game is
simply an interactive proof system in which the verifier's messages to the
prover consist only of uniformly generated bits from the public random source.

Although Arthur-Merlin games are more restricted than interactive proof
systems in the sense just described, the two models are known to be
computationally equivalent.
In particular, any language having an interactive proof system in which
a constant number of messages is exchanged between the prover and verifier
also has an Arthur-Merlin game in which precisely two messages are
exchanged, the first from Arthur to Merlin and the second from Merlin back to
Arthur \cite{GoldwasserS89, BabaiM88}.
The complexity class consisting of all such languages is $\mathrm{AM}$.
Also following from \cite{GoldwasserS89} is the fact that any language having
an unrestricted (polynomial-message) interactive proof system also has a
polynomial-message Arthur-Merlin game.
The complexity class consisting of all such languages was initially called
$\mathrm{IP}$, but is now known to be equal to $\mathrm{PSPACE}$
\cite{LundF+92, Shamir92}.

A third complexity class arising from these models is $\mathrm{MA}$, which is
the class consisting of all languages having an interactive proof system in
which a single message is sent, from the prover to the verifier.
One may view the definition of this class as a slight variation on the
``guess and check'' definition of $\mathrm{NP}$, where instead of being
deterministic the checking procedure may use randomness.
As the usual convention for Arthur-Merlin games is to disallow Arthur the use
of the public random source except for the generation of messages, the class
$\mathrm{MA}$ would typically be described as consisting of all languages
having two-message Arthur-Merlin games in which the first message is sent from
Merlin to Arthur and the second from Arthur to Merlin.
However, given that the information transmitted to Merlin in the second
message is irrelevant from the point of view of the game, and may instead be
viewed as just a use of the random source and not as a message, it is natural
to refer to such games as one-message Arthur-Merlin games.

Quantum computational variants of interactive proof systems have previously
been considered in several papers, including the general multiple-message case
\cite{Watrous03-pspace, KitaevW00, KobayashiM03, RosgenW04, GutoskiW05}
as well as the single-message case
\cite{AharonovR03,JanzingW+03, KempeR03, KempeK+04, KobayashiM+03,
RazS04,Vyalyi03,Watrous00}.
As for classical interactive proof systems, quantum interactive proof systems
consist of two parties---a prover with unlimited computation power and a
computationally bounded verifier.
Now, however, the two parties may process and exchange quantum information.
The complexity class consisting of all languages having quantum interactive
proof systems is denoted $\mathrm{QIP}$, and satisfies
$\mathrm{PSPACE}\subseteq\mathrm{QIP}\subseteq\mathrm{EXP}$ \cite{KitaevW00}.
Here, $\mathrm{EXP}$ denotes the class of languages decidable by a
deterministic Turing machine running in time $2^{q}$ for some polynomial $q$.

There are both similarities and some apparent differences in the properties of
quantum and classical interactive proof systems.
Perhaps the most significant difference is that any language having an
unrestricted (polynomial-message) quantum interactive proof system also has a
three-message quantum interactive proof system \cite{KitaevW00}.
This cannot happen classically unless $\mathrm{AM} = \mathrm{PSPACE}$.

This paper investigates various aspects of quantum Arthur-Merlin games.
In analogy to the classical case, we define quantum Arthur-Merlin games
to be restricted forms of quantum interactive proof systems in which the
verifier's (Arthur's) messages to the prover (Merlin) are uniformly generated
random bits, as opposed to arbitrary messages.
Consequently, Arthur is not capable of sending quantum information to Merlin
at any point during a quantum Arthur-Merlin game.
Similar to the classical case, quantum Arthur-Merlin games give rise to
complexity classes depending on the number of messages exchanged
between Arthur and Merlin.
In particular, we obtain three primary complexity classes corresponding to
Arthur-Merlin games with one message, two messages, and three or more messages.

In the one-message case, Merlin sends a single message to Arthur, who checks
it and makes a decision to accept or reject the input.
The corresponding complexity class is denoted $\mathrm{QMA}$, and has
been considered previously in the papers cited above.
In this situation Merlin's message to Arthur may simply be viewed as a quantum
witness or certificate that Arthur checks in polynomial time with a quantum
computer.
To our knowledge, the idea of a quantum state playing the role of a certificate
in this sense was first proposed by \cite{Knill96}, and the idea was
later studied in greater depth by \cite{Kitaev99}.
Kitaev proved various fundamental properties of $\mathrm{QMA}$, which are
described in \cite{KitaevS+02} and \cite{AharonovN02}.

One of the facts that Kitaev proved was that the completeness and soundness
errors in a $\mathrm{QMA}$ protocol may be efficiently reduced by parallel
repetition.
Because quantum information cannot be copied, however, and Arthur's
verification procedure is potentially destructive to Merlin's message,
Arthur requires multiple copies of Merlin's message for this method to work.
This method therefore requires a polynomial increase in the length of
Merlin's message to Arthur in order to achieve exponentially decreasing error.
In this paper, we prove that this increase in the length of Merlin's message
is not required after all---using a different error reduction method, an
exponential reduction in error is possible with no increase whatsoever in the
length of Merlin's message to Arthur.

It is known that $\mathrm{QMA}$ is contained in the class $\mathrm{PP}$, which
can be proved using the $\mathrm{GapP}$-based method of
\cite{FortnowR99} together with some simple facts from matrix analysis.
This fact was noted without proof in \cite{KitaevW00}.
A proof of this fact was, however, given by \cite{Vyalyi03}, who in fact
strengthened this result to show that $\mathrm{QMA}$ is contained in a
subclass $\mathrm{A}_0\mathrm{PP}$ of $\mathrm{PP}$.
(Definitions of the classes $\mathrm{PP}$ and $\mathrm{A}_0\mathrm{PP}$
can be found in \ref{sec:preliminaries} of this paper.)
Based on our new error reduction method, we give a simplified proof of this
containment.
We also use our error reduction method to prove that one-message quantum
Arthur-Merlin games in which Merlin's message has logarithmic length give no
increase in power over~$\mathrm{BQP}$.

In the two-message case, Arthur flips some number of fair coins, sends the
results of those coin-flips to Merlin, and Merlin responds with some quantum
state.
Arthur performs a polynomial-time quantum computation on the random bits
together with Merlin's response, which determines whether Arthur accepts or
rejects.
The corresponding complexity class will be denoted $\mathrm{QAM}$.
Two facts about $\mathrm{QAM}$ are proved in this paper.
The first is the very basic fact that parallel repetition reduces error
exactly as in the classical case.
(This fact does not follow from known facts about quantum interactive proof
systems, as parallel repetition is only known to reduce error for general
quantum interactive proof systems having perfect completeness.)
The second fact is that $\mathrm{QAM}$ is contained in
$\mathrm{BP}\cdot\mathrm{PP}$, the class obtained by applying the $\mathrm{BP}$
operator to the class $\mathrm{PP}$.

Finally, in the three-message case, Merlin sends Arthur a message consisting
of some number of qubits, Arthur flips some number of fair coins and sends the
results to Merlin, and then Merlin responds with a second collection of
qubits.
Arthur performs a polynomial-time quantum computation on all of the
qubits sent by Merlin together with the values of his own coin-flips,
and decides whether to accept or reject.
The corresponding complexity class will be denoted $\mathrm{QMAM}$.
It is proved that any language having an ordinary quantum interactive
proof system is contained in $\mathrm{QMAM}$, implying
$\mathrm{QMAM} = \mathrm{QIP}$.

In spirit, the equality $\mathrm{QMAM} = \mathrm{QIP}$ resembles the theorem 
of \cite{GoldwasserS89} establishing that classical Arthur-Merlin games
and interactive proof systems are equivalent in power.
However, there is no similarity in the proofs of these facts.
Moreover, our result is stronger than what is likely to hold classically.
Specifically, we prove that any language having a quantum interactive proof
system also has a three-message quantum Arthur-Merlin game in which Arthur's
only message to Merlin consists of just a single coin-flip (in order to
achieve perfect completeness and soundness error exponentially close to 1/2).
This is impossible classically unless interaction is useless in classical
interactive proof systems; for if Arthur flips only one coin, Merlin may as
well send his first message and the two possible second messages to Arthur in
a single message.
The reason why this strategy fails in the quantum case is that Merlin's
first and second messages may need to be entangled in order to be convincing
to Arthur, but it may not possible for Merlin to simultaneously entangle his
two possible second messages with the first in a way that convinces Arthur to
accept.
This is an example of the principle that Bennett refers to as the ``monogamy of
entanglement'' (see, for example, \cite{Terhal04}); the more a given system
is entangled with a second system, the less it can be entangled with a third.

\subsection*{Organization of the paper}

The remainder of this paper is organized as follows.
We begin with \ref{sec:preliminaries}, which discusses background information
needed elsewhere in the paper, including a summary of basic notation
and conventions that are used, definitions of some relevant counting
complexity classes,
and background on quantum computation and quantum interactive proof systems.
The next three sections correspond to the three complexity classes
$\mathrm{QMA}$, $\mathrm{QAM}$, and $\mathrm{QMAM}$, respectively;
\ref{sec:QMA} discusses one-message quantum Arthur-Merlin games,
\ref{sec:QAM} discusses the two-message case, and
\ref{sec:QMAM} discusses the case of three or more messages.
The paper concludes with \ref{sec:conclusion}, which mentions
some open problems relating to quantum Arthur-Merlin games.


\section{Background Information}
\label{sec:preliminaries}

This section summarizes various background information that is needed for
the remainder of the paper, including information on quantum computation,
counting complexity, and quantum interactive proof systems.

We begin with some remarks about notation and other simple conventions that
are followed throughout.
All strings and languages in this paper will be over the alphabet
$\Sigma = \{0,1\}$.
We denote by $\mathit{poly}$ the set of all functions
$f:\mathbb{N}\rightarrow\mathbb{N}\backslash\{0\}$ 
(where $\mathbb{N} = \{0,1,2,\ldots\}$) for which there exists a
polynomial-time deterministic Turing machine that outputs $1^{f(n)}$ on
input $1^n$.
For every integer $k\geq 2$, we fix a polynomial-time computable function 
that, for every choice of $x_1,\ldots,x_k\in\Sigma^{\ast}$, encodes the
$k$-tuple $(x_1,\ldots,x_k)$ as a single element of $\Sigma^{\ast}$.
These functions are assumed to satisfy the usual properties of tuple-functions,
namely that they are one-to-one and polynomial-time invertible in each
argument.
As is typical, reference to these functions is often implicit;
for instance, we write $f(x_1,\ldots,x_k)$ as shorthand for
$f((x_1,\ldots,x_k))$ when $x_1,\ldots,x_k\in\Sigma^{\ast}$ and the domain
of the function $f$ is understood to be $\Sigma^{\ast}$.

\subsection*{Quantum computation}
We will assume that the reader has familiarity with the mathematics of
quantum information, which is discussed in the books of \cite{KitaevS+02}
and \cite{NielsenC00}.
The quantum complexity classes discussed in this paper are based on the
quantum circuit model, with which we also assume familiarity.

All quantum circuits considered in this paper will be assumed to be
composed only of Toffoli gates, Hadamard gates, and $i$-shift gates
(which induce the mapping $\ket{0}\mapsto\ket{0}$, $\ket{1}\mapsto i\ket{1}$).
This is a universal set of gates \cite{Kitaev97}, so there is no loss of
generality in restricting our attention to this set.
We assume that a reasonable encoding scheme has been fixed that allows quantum
circuits to be encoded as binary strings having length at least the size of
the encoded circuit and at most some fixed polynomial in the circuit's size.

A collection $\{A_x:x\in\Sigma^{\ast}\}$ of quantum circuits is said to be 
{\em generated in polynomial-time} if there exists a polynomial-time
deterministic Turing machine that, on input $x\in\Sigma^{\ast}$, outputs
an encoding of the circuit $A_x$.
When such a family is parameterized by tuples of strings, it is to be
understood that we are implicitly referring to one of the tuple-functions
discussed previously.
For instance, we will consider families of the form
$\{A_{x,y}\,:\,x,y\in\Sigma^{\ast}\}$ when two- and three-message quantum
Arthur-Merlin games are discussed.

The notion of a polynomial-time generated family is similar to the usual
notion of a polynomial-time uniform family of circuits, except that it allows
the procedure generating the circuits to have access to the input $x$
rather than just the length of $x$ written in unary.
In essence, the input $x$ may be ``hard-coded'' into a given circuit
in a polynomial-time generated family, so that it is not necessary to
assume that the input $x$ is input to the circuit itself.
This is simply done as a matter of convenience and simplicity---all of
the polynomial-time generated families of quantum circuits in this paper
could be replaced by polynomial-time uniform families where the string
given to the generating procedure is instead input directly into the circuit.

Let us illustrate the use of polynomial-time generated families of quantum
circuits by defining $\mathrm{BQP}$, the class of languages recognizable
in quantum polynomial time with bounded error.
A language $L$ is in $\mathrm{BQP}$ if and only if there exists a
polynomial-time generated family $\{A_x\}$ of quantum circuits
such that the following conditions hold.
First, it is required that there exist a function $k\in\mathit{poly}$ 
such that each circuit $A_x$ act on precisely $k(|x|)$ qubits.
(This condition is not really necessary, but will simplify further
discussions.)
Let $\Pi_1 = \ket{1}\bra{1}\otimes I_{k-1}$,
where $k$ is shorthand for $k(|x|)$ and, in general,
$I_n$ denotes the identity operator acting on $n$ qubits.
Then it is required that
\begin{mylist}{\parindent}
\item[1.]
if $x\in L$ then
$\left\|\Pi_1 A_x \ket{0^k}\right\|^2 \geq \frac{2}{3}$,
and
\item[2.]
if $x\not\in L$ then
$\left\|\Pi_1 A_x \ket{0^k}\right\|^2 \leq\frac{1}{3}$.
\end{mylist}
In words, if the input is $x$, then the circuit $A_x$ is run on the
all-zero input and the first qubit is measured in the standard basis.
If the measurement result is~1, the computation is viewed as accepting,
otherwise it is rejecting.
The usual notion of bounded error is required.

It will sometimes be helpful when describing certain quantum Arthur-Merlin
games to refer to {\em quantum registers}.
These are simply collections of qubits to which we assign some name.
When we refer to the {\em reduced state} of a given register, we mean
the mixed state obtained by tracing out all other registers beside
the one to which we are referring.

\subsection*{Counting classes}
Some of the results in this paper involve relations between complexity
classes based on quantum Arthur-Merlin games and classes based on the
notion of counting complexity.
Here we briefly discuss this notion and the classes relevant to this paper;
for more information about counting complexity, see \cite{Fortnow97}.

A function $f:\Sigma^{\ast}\rightarrow\mathbb{N}$ is an element of the
function class $\mathrm{\#P}$ if and only if there exists a polynomial-time
nondeterministic Turing machine that, on each input $x\in\Sigma^{\ast}$,
has precisely $f(x)$ accepting computation paths.
For any function $f\in\mathrm{\#P}$ there exists a function $q\in\mathit{poly}$
such that $f(x) \leq 2^{q(|x|)}$ for all $x\in\Sigma^{\ast}$.

A function $f:\Sigma^{\ast}\rightarrow\mathbb{Z}$ is an element of the
function class $\mathrm{FP}$ if it is computable in polynomial time,
with the understanding that the output of the function is the integer
represented in binary notation by the output of the computation.

A function $f:\Sigma^{\ast}\rightarrow\mathbb{Z}$ is an element of the
function class $\mathrm{GapP}$ if and only if there exist functions
$g,h\in\mathrm{\#P}$ such that $f(x) = g(x) - h(x)$ for all
$x\in\Sigma^{\ast}$.
The function class $\mathrm{GapP}$ possesses remarkable closure
properties, including closure under subtraction, exponential sums, and
polynomial products.
In particular, if $f\in\mathrm{GapP}$ and $q\in\mathit{poly}$, then the
functions $g$ and $h$ defined as
\[
g(x) = \sum_{i=1}^{2^{q(|x|)}} f(x,i),\quad
h(x) = \prod_{i = 1}^{q(|x|)} f(x,i)
\]
are elements of $\mathrm{GapP}$.
(Here the integer $i$ is identified with the string having no leading zeroes
that encodes it in binary notation.)
It is not difficult to show that $\mathrm{FP}\subseteq\mathrm{GapP}$.

The complexity class $\mathrm{PP}$ consists of all languages
$L\subseteq\Sigma^{\ast}$ for which there exists a function
$f\in\mathrm{GapP}$ such that  $x\in L$ if and only if $f(x) > 0$
for all $x\in\Sigma^{\ast}$.
The class $\mathrm{A}_0\mathrm{PP}$ consists of all languages
$L\subseteq\Sigma^{\ast}$ for which there exist functions
$f\in\mathrm{GapP}$ and $g\in\mathrm{FP}$ satisfying
\[
x\in L \;\Rightarrow\;f(x) \geq g(x),\quad
x\not\in L \;\Rightarrow\;0 \leq f(x)\leq \frac{g(x)}{2},
\]
for all $x\in\Sigma^{\ast}$.
Finally, the complexity class $\mathrm{BP}\cdot\mathrm{PP}$ refers to the
$\mathrm{BP}$ operator applied to the class $\mathrm{PP}$;
it contains all languages $L\subseteq\Sigma^{\ast}$ such that there
exists a language $A\in\mathrm{PP}$ and a function $q\in\mathit{poly}$
such that
\[
\left|\left\{ y\in\Sigma^{q(|x|)}\,:\,(x,y) \in A
\;\Leftrightarrow\; x\in L\right\}\right|
\geq \frac{2}{3}\,2^{q(|x|)}.
\]

Counting complexity and quantum complexity were related by \cite{FortnowR99},
who gave a simple proof that $\mathrm{BQP}\subseteq\mathrm{PP}$ based
on the closure properties of $\mathrm{GapP}$ functions discussed above.
(The containment $\mathrm{BQP}\subseteq\mathrm{PP}$ had been proved
earlier by \cite{AdlemanD+97} using a different method.)
In fact, Fortnow \& Rogers proved the stronger containment
$\mathrm{BQP}\subseteq\mathrm{AWPP}$, where $\mathrm{AWPP}$ is a
subclass of $\mathrm{PP}$ that we will not define in this paper.
As a couple of the facts we prove are based on the method of
Fortnow \& Rogers, it will be helpful for us to summarize this method.
The quantum Turing machine model was used in the original proof, but our
summary is instead based on polynomial-time generated families of quantum
circuits.

Suppose that $L\in\mathrm{BQP}$, which implies the existence of a
polynomial-time generated family $\{A_x\}$ of quantum circuits satisfying
the conditions of the definition of $\mathrm{BQP}$ discussed previously.
The goal is to construct a $\mathrm{GapP}$ function $f$ and a
polynomially bounded $\mathrm{FP}$ function $g$ such that
\[
\frac{f(x)}{2^{g(x)}} = \bra{0^k} A_x^{\dagger} \Pi_1 A_x \ket{0^k}
= \left\|\Pi_1 A_x \ket{0^k}\right\|^2.
\]
Once this is done, the $\mathrm{GapP}$ function $h(x) = 2 f(x) - 2^{g(x)}$
satisfies the required property to establish $L\in\mathrm{PP}$; namely that
$h(x) > 0$ if and only if $x\in L$.

The functions $f$ and $g$ are of course based on the circuit family $\{A_x\}$.
For a given string $x$, assume that the circuit $A_x$ consists of gates
$G_1,\ldots,G_{q(|x|)}$ for some function $q\in\mathit{poly}$.
Each of the gates $G_j$, when tensored with the identity operator on
the qubits not affected by $G_j$, gives rise to a $2^k \times 2^k$
matrix whose individual entries, indexed by pairs of strings of length $k$,
can be computed in polynomial time given $x$.
These entries are elements of the set
\[
\left\{0,1,i,1/\sqrt{2},-1/\sqrt{2}\right\}
\]
because we assume $A_x$ is composed only of Toffoli, Hadamard, and
$i$-shift gates.
Similarly, $\Pi_1$ is a $2^k \times 2^k$ matrix whose entries (this time
restricted to the set $\{0,1\}$) are also computable in polynomial time
given $x$.

The value $\bra{0^k} A_x^{\dagger} \Pi_1 A_x \ket{0^k}$ therefore corresponds
to the $(0^k,0^k)$ entry of the matrix product
\[
G_1^{\dagger}\cdots G_q^{\dagger}\,\Pi_1\, G_q\cdots G_1,
\]
which can be expressed as an exponential sum of a polynomial product of the
entries of these matrices.
By letting the function $g$ represents the total number of Hadamard transforms
in the circuit $A_x$, it is fairly straightforward to construct an
appropriate $\mathrm{GapP}$ function $f$ based on closure properties of
the class $\mathrm{GapP}$.
Further details can be found in \cite{FortnowR99} as well as in
\cite{Vyalyi03}.


\subsection*{Quantum interactive proofs}
\label{sec:qip}

Here we discuss background information on quantum interactive proof systems
that will be used later in the paper when it is proved that quantum
Arthur-Merlin games have the same power as arbitrary quantum interactive
proof systems.
It will only be necessary for us to discuss the particular case of
three-message quantum interactive proof systems, as any polynomial-message
quantum interactive proof system can be simulated by a three-message
quantum interactive proof.
Moreover, such a proof system may be taken to have perfect completeness and
exponentially small soundness error.
These facts are proved in \cite{KitaevW00}, to which the reader is referred
for a more complete discussion of quantum interactive proof systems.

For a fixed input~$x$, a three-message quantum interactive proof system
operates as follows.
The verifier begins with a $k$-qubit register $\mathsf{V}$ and the prover
begins with two registers: an $m$-qubit register $\mathsf{M}$ and
an $l$-qubit register $\mathsf{P}$.
The register $\mathsf{V}$ corresponds to the verifier's work-space,
the register $\mathsf{M}$ corresponds to the message qubits that
are sent back and forth between the prover and verifier, and the register
$\mathsf{P}$ corresponds to the prover's workspace.
The register $\mathsf{M}$ begins in the prover's possession because
the prover sends the first message.
The verifier's work-space register $\mathsf{V}$ begins initialized to the
state $\ket{0^k}$, while the prover initializes the pair
$(\mathsf{M},\mathsf{P})$ to some arbitrary quantum state $\ket{\psi}$.

In the first message, the prover sends $\mathsf{M}$ to the verifier.
The verifier applies some unitary transformation $V_1$ to the pair 
$(\mathsf{V},\mathsf{M})$ and returns $\mathsf{M}$ to the prover in the
second message.
The prover now applies some arbitrary unitary transformation $U$ to the pair
$(\mathsf{M},\mathsf{P})$ and returns $\mathsf{M}$ to the verifier in the
third and final message.
Finally, the verifier applies a second unitary transformation $V_2$ to
the pair $(\mathsf{V},\mathsf{M})$ and measures the first qubit
of the resulting collection of qubits in the standard basis.
The outcome 1 is interpreted as ``accept'' and 0 is interpreted as ``reject''.

Let $\Pi_0$, $\Pi_1$, $\Delta_0$, and $\Delta_1$ be projections defined as
\[
\Pi_1 = \ket{1}\bra{1}\otimes I_{k+m-1},\quad
\Delta_1 = \ket{0^k}\bra{0^k}\otimes I_m,\quad
\Pi_0 = \ket{0}\bra{0}\otimes I_{k+m-1},\quad
\Delta_0 = I_{k+m} - \Delta_1.
\]
In other words, these are $k+m$ qubit projections that act on the pair of
registers $(\mathsf{V},\mathsf{M})$;
$\Pi_1$ and $\Pi_0$ are projections onto those states for which the first
qubit of the register $\mathsf{V}$ is 1 or 0, respectively, and $\Delta_1$
and $\Delta_0$ are projections onto those states for which the register
$\mathsf{V}$ contains the state $\ket{0^k}$ or contains a state orthogonal
to $\ket{0^k}$, respectively.

The maximum probability with which a verifier specified by $V_1$ and $V_2$
can be made to accept is
\begin{equation}\label{eq:max-accept}
\left\| (\Pi_1V_2\otimes I_l) (I_k\otimes U) (V_1\otimes I_l)
(\ket{0^k}\ket{\psi})\right\|^2,
\end{equation}
maximized over all choices of the state $\ket{\psi}$ and the unitary
transformation $U$.
The number $l$ is determined by the prover's strategy, so one may maximize
over this number as well.
However, there is no loss of generality in assuming $l = m+k$;
with this many work qubits, the prover may store a purification of
the reduced state of the pair $(\mathsf{V},\mathsf{M})$, which is sufficient
for an optimal strategy.

There is another way to characterize the maximum acceptance probability
for a given verifier based on the fidelity function
\[
F(\rho,\xi)\:=\:\op{tr}\sqrt{\sqrt{\rho}\,\xi\sqrt{\rho}}.
\]
To describe this characterization we will need to define
various sets of states of the pair of registers $(\mathsf{V},\mathsf{M})$.
For any projection $\Lambda$ on $k+m$ qubits let $\mathcal{S}(\Lambda)$ denote
the set of all mixed states $\rho$ of $(\mathsf{V},\mathsf{M})$ that satisfy
$\rho = \Lambda \rho \Lambda$, i.e., the collection of states whose support
is contained in the space onto which $\Lambda$ projects.
Also let $\mathcal{S}_{\mathsf{V}}(\Lambda)$ denote the set of all reduced
states of $\mathsf{V}$ that result from some state
$\rho\in\mathcal{S}(\Lambda)$, i.e.,
\[
\mathcal{S}_{\mathsf{V}}(\Lambda) =
\left\{\op{tr}_{\mathsf{M}}\rho\,:\,\rho\in\mathcal{S}(\Lambda)\right\},
\]
where $\op{tr}_{\mathsf{M}}$ denotes the partial trace over the register
$\mathsf{M}$.

\begin{prop}\label{prop:max-accept}
The maximum probability with which a verifier specified by
$V_1$ and $V_2$ can be made to accept is
\[
\max\left\{ F(\rho,\xi)^2\,:\,
\rho\in \mathcal{S}_{\mathsf{V}}(V_1 \Delta_1 V_1^{\dagger}),\;
\xi \in \mathcal{S}_{\mathsf{V}}(V_2^{\dagger} \Pi_1 V_2)\right\}.
\]
\end{prop}

\noindent
This proposition is essentially a restatement based on
Uhlmann's Theorem (see \cite{NielsenC00}), of the fact that the
quantity \ref{eq:max-accept} above represents the maximum acceptance
probability of the verifier described by $V_1$ and~$V_2$.
This equivalence is discussed further in \cite{KitaevW00}.


\section{QMA}
\label{sec:QMA}


A $\mathrm{QMA}$ verification procedure $A$ is a family
of quantum circuits $\{A_x:x\in\Sigma^{\ast}\}$ that is generated in
polynomial time, together with a function $m\in\mathit{poly}$.
The function $m$ specifies the length of Merlin's message to Arthur, and
it is assumed that each circuit $A_x$ acts on $m(|x|) + k(|x|)$ qubits for
some function $k$ specifying the number of work qubits used by the circuit.
As we have done in the previous section, when the input $x$ has been fixed or
is implicit we will generally write $m$ to mean $m(|x|)$, $k$ to mean $k(|x|)$,
and so forth, in order to simplify our notation.
When we want to emphasize the length of Merlin's message, we will refer to $A$
as an $m$-qubit $\mathrm{QMA}$ verification procedure.

Consider the following process for a string $x\in\Sigma^{\ast}$ and a quantum
state $\ket{\psi}$ on $m$ qubits:
\begin{mylist}{\parindent}
\item[1.]
Run the circuit $A_x$ on the input state $\ket{\psi}\ket{0^k}$.
\item[2.]
Measure the first qubit of the resulting state in the standard basis,
interpreting the outcome 1 as {\em accept} and the outcome 0 as {\em reject}.
\end{mylist}
The probability associated with the two possible outcomes
will be referred to as
$\op{Pr}[\mbox{$A_x$ accepts $\ket{\psi}$}]$ and
$\op{Pr}[\mbox{$A_x$ rejects $\ket{\psi}$}]$ accordingly.

\begin{definition}
\label{def:QMA}
The class $\mathrm{QMA}(a,b)$ consists of all languages
$L\subseteq \Sigma^{\ast}$ for which there exists a $\mathrm{QMA}$ verification
procedure $A$
for which the following holds:
\begin{mylist}{\parindent}
\item[1.]
For all $x\in L$ there exists an $m$ qubit quantum state $\ket{\psi}$
such that $\op{Pr}[\mbox{$A_x$ accepts $\ket{\psi}$}]\geq a$.
\item[2.] 
For all $x\not\in L$ and all $m$ qubit quantum states $\ket{\psi}$,
$\op{Pr}[\mbox{$A_x$ accepts $\ket{\psi}$}]\leq b$.
\end{mylist}
For any $m\in\mathit{poly}$, the class
$\mathrm{QMA}_m(a,b)$ consists of all languages $L\subseteq\Sigma^{\ast}$
for which there exists an $m$-qubit $\mathrm{QMA}$ verification procedure that
satisfies the above properties.
\end{definition}

One may consider the cases where $a$ and $b$ are constants or functions
of the input length $n = |x|$ in this definition.
If $a$ and $b$ are functions of the input length, it is assumed that $a(n)$
and $b(n)$ can be computed deterministically in time polynomial in $n$.
When no reference is made to the probabilities $a$ and $b$, it is assumed
$a = 2/3$ and $b = 1/3$.


\subsection*{Strong Error Reduction}

It is known that $\mathrm{QMA}$ is robust with respect to error bounds in the
following sense.
\begin{theorem}[Kitaev]
\label{theorem:Kitaev1}
Let $a,b:\mathbb{N}\rightarrow [0,1]$ and $q\in\mathit{poly}$ satisfy
\[
a(n) - b(n) \geq \frac{1}{q(n)}
\]
for all $n\in\mathbb{N}$.
Then $\mathrm{QMA}(a,b) \subseteq \mathrm{QMA}(1 - 2^{-r},2^{-r})$
for every $r\in\mathit{poly}$.
\end{theorem}

A proof of this theorem appears in Section 14.2 of \cite{KitaevS+02}.
The idea of the proof is as follows.
If we have a verification procedure $A$ with completeness and soundness
probabilities given by $a$ and $b$, we construct a new verification procedure
that independently runs $A$ on some sufficiently large number of copies
of the original certificate and accepts if the number of acceptances of $A$
is larger than $(a+b)/2$.
The only difficulty in proving that this construction works lies in the fact
that the new certificate cannot be assumed to consist of several copies of the
original certificate, but may be an arbitrary (possibly highly entangled)
quantum state.
Intuitively, however, entanglement cannot help Merlin to cheat; under the
assumption that $x\not\in L$, the probability of acceptance for any particular
execution of $A$ is bounded above by $b$, and this is true regardless of
whether one conditions on the outcomes of any of the other executions of~$A$.
This construction requires an increase in the length of Merlin's message
to Arthur in order to reduce error.

The main result of this section is the following theorem, which states
that one may decrease error without any increase in the length of Merlin's
message.

\begin{theorem}
\label{theorem:amplify}
Let $a,b:\mathbb{N}\rightarrow [0,1]$ and $q\in\mathit{poly}$ satisfy
\[
a(n) - b(n) \geq \frac{1}{q(n)}
\]
for all $n\in\mathbb{N}$.
Then $\mathrm{QMA}_m(a,b) \subseteq \mathrm{QMA}_m(1 - 2^{-r},2^{-r})$
for every $m,r\in\mathit{poly}$.
\end{theorem}

\begin{proof}
Assume $L\in\mathrm{QMA}_m(a,b)$, and $A$ is an $m$-qubit $\mathrm{QMA}$
verification procedure that witnesses this fact.
We will describe a new $m$-qubit $\mathrm{QMA}$ verification procedure $B$ with
exponentially small completeness and soundness error for the language $L$,
which will suffice to prove the theorem.

It will simplify matters to assume hereafter that the input $x$ is
fixed---it will be clear that the new verification procedure can
be generated in polynomial-time.
As the input $x$ is fixed, we will write $A$ and $B$ to denote $A_x$
and $B_x$, respectively.

It will be helpful to refer to the $m$ message qubits along with the $k$
work-space qubits of $A$ as a single $m+k$ qubit quantum register $\mathsf{R}$.
Define projections acting on the vector space corresponding to $\mathsf{R}$
as follows:
\begin{equation}\label{eq:projections}
\Pi_1 = \ket{1}\bra{1}\otimes I_{m+k-1}, \quad
\Delta_1 = I_m\otimes\ket{0^k}\bra{0^k}, \quad
\Pi_0 = \ket{0}\bra{0}\otimes I_{m+k-1}, \quad
\Delta_0 = I_{m+k} - \Delta_1.
\end{equation}
The measurement described by $\{\Pi_0,\Pi_1\}$ is just a measurement of the
first qubit of $\mathsf{R}$ in the computational basis; this measurement
determines whether Arthur accepts or rejects after the circuit $A$ is
applied.
The measurement described by $\{\Delta_0,\Delta_1\}$ gives outcome 1 if the
last $k$ qubits of $\mathsf{R}$, which correspond to Arthur's work-space
qubits, are set to their initial all-zero state, and gives outcome 0 otherwise.
(These projections are similar to those in \ref{sec:preliminaries} except
that the message qubits and Arthur's work qubits are reversed for notational
convenience.)

The procedure $B$ operates as follows.
It assumes that initially the first $m$ qubits of $\mathsf{R}$ contain
Merlin's message $\ket{\psi}$ and the remaining $k$ qubits are set to the
state $\ket{0^k}$.

\begin{mylist}{\parindent}
\item[1.]
Set $y_0 \leftarrow 1$ and $i\leftarrow 1$.
\vspace{1mm}

\item[2.]
Repeat:
\begin{mylist}{7mm}
\item[a.]
Apply $A$ to $\mathsf{R}$ and measure $\mathsf{R}$ with respect to the
measurement described by $\{\Pi_0,\Pi_1\}$.
Let $y_i$ denote the outcome, and set $i\leftarrow i+1$.
\vspace{1mm}

\item[b.]
Apply $A^{\dagger}$ to $\mathsf{R}$ and measure $\mathsf{R}$ with respect
to the measurement described by $\{\Delta_0,\Delta_1\}$.
Let $y_i$ denote the outcome, and set $i\leftarrow i+1$.
\end{mylist}

\noindent
Until $i\geq N$, where $N = 8\,q^2 r$.
\vspace{1mm}

\item[3.]
For each $i=1,\ldots,N$ set
\[
z_i \leftarrow \left\{\begin{array}{ll}1 & \mbox{if $y_i = y_{i-1}$}\\
0 & \mbox{if $y_i \not= y_{i-1}$.}\end{array}\right.
\]
\noindent
Accept if $\sum_{i=1}^N z_i \geq N\cdot\frac{a+b}{2}$ and reject otherwise.
\end{mylist}
Although the description of this procedure refers to various
measurements, it is possible to simulate these measurements with unitary
gates in the standard way, which allows the entire procedure to be implemented
by a unitary quantum circuit.
\ref{fig:circuit} illustrates a quantum circuit implementing
this procedure for the case $N=5$.
\begin{figure*}[t]
\begin{center}
\setlength{\unitlength}{1910sp}
\begin{picture}(11424,3999)(800,-4700)
\put(1,-1950){\makebox(0,0)[lb]{$\ket{\psi}\;\left\{
\rule{0mm}{6mm}\right.$}}
\put(1201,-1111){\line( 1, 0){600}}
\put(1201,-1261){\line( 1, 0){600}}
\put(1201,-1411){\line( 1, 0){600}}
\put(1201,-1561){\line( 1, 0){600}}
\put(1201,-1711){\line( 1, 0){600}}
\put(1201,-2911){\line( 1, 0){600}}
\put(1201,-2161){\line( 1, 0){600}}
\put(1201,-2311){\line( 1, 0){600}}
\put(1201,-2461){\line( 1, 0){600}}
\put(1201,-2611){\line( 1, 0){600}}
\put(1201,-2761){\line( 1, 0){600}}
\put(1201,-1861){\line( 1, 0){600}}
\put(3001,-1111){\line( 1, 0){600}}
\put(3001,-1411){\line( 1, 0){600}}
\put(3001,-1561){\line( 1, 0){600}}
\put(3001,-1711){\line( 1, 0){600}}
\put(3001,-1861){\line( 1, 0){600}}
\put(3001,-2011){\line( 1, 0){600}}
\put(3001,-2161){\line( 1, 0){600}}
\put(3001,-2311){\line( 1, 0){600}}
\put(3001,-2461){\line( 1, 0){600}}
\put(3001,-2611){\line( 1, 0){600}}
\put(3001,-2761){\line( 1, 0){600}}
\put(3001,-2911){\line( 1, 0){600}}
\put(4801,-1111){\line( 1, 0){600}}
\put(4801,-1261){\line( 1, 0){600}}
\put(4801,-1411){\line( 1, 0){600}}
\put(4801,-1561){\line( 1, 0){600}}
\put(4801,-1711){\line( 1, 0){600}}
\put(4801,-1861){\line( 1, 0){600}}
\put(4801,-2161){\line( 1, 0){220}}
\put(4801,-2311){\line( 1, 0){220}}
\put(4801,-2461){\line( 1, 0){220}}
\put(4801,-2611){\line( 1, 0){220}}
\put(4801,-2761){\line( 1, 0){220}}
\put(4801,-2911){\line( 1, 0){220}}
\put(5181,-2161){\line( 1, 0){220}}
\put(5181,-2311){\line( 1, 0){220}}
\put(5181,-2461){\line( 1, 0){220}}
\put(5181,-2611){\line( 1, 0){220}}
\put(5181,-2761){\line( 1, 0){220}}
\put(5181,-2911){\line( 1, 0){220}}
\put(6601,-1111){\line( 1, 0){600}}
\put(6601,-1411){\line( 1, 0){600}}
\put(6601,-1561){\line( 1, 0){600}}
\put(6601,-1711){\line( 1, 0){600}}
\put(6601,-1861){\line( 1, 0){600}}
\put(6601,-2011){\line( 1, 0){600}}
\put(6601,-2161){\line( 1, 0){600}}
\put(6601,-2311){\line( 1, 0){600}}
\put(6601,-2461){\line( 1, 0){600}}
\put(6601,-2611){\line( 1, 0){600}}
\put(6601,-2761){\line( 1, 0){600}}
\put(6601,-2911){\line( 1, 0){600}}
\put(8401,-1111){\line( 1, 0){600}}
\put(8401,-1261){\line( 1, 0){600}}
\put(8401,-1411){\line( 1, 0){600}}
\put(8401,-1561){\line( 1, 0){600}}
\put(8401,-1711){\line( 1, 0){600}}
\put(8401,-1861){\line( 1, 0){600}}
\put(8401,-2161){\line( 1, 0){220}}
\put(8781,-2161){\line( 1, 0){220}}
\put(8401,-2311){\line( 1, 0){220}}
\put(8781,-2311){\line( 1, 0){220}}
\put(8401,-2461){\line( 1, 0){220}}
\put(8781,-2461){\line( 1, 0){220}}
\put(8401,-2611){\line( 1, 0){220}}
\put(8781,-2611){\line( 1, 0){220}}
\put(8401,-2761){\line( 1, 0){220}}
\put(8781,-2761){\line( 1, 0){220}}
\put(8401,-2911){\line( 1, 0){220}}
\put(8781,-2911){\line( 1, 0){220}}
\put(3301,-1111){\circle*{135}}
\put(6901,-1111){\circle*{135}}
\put(10501,-1111){\circle*{135}}
\put(3301,-3961){\circle{150}}
\put(5101,-4111){\circle{150}}
\put(6901,-4261){\circle{150}}
\put(8701,-4411){\circle{150}}
\put(10501,-4561){\circle{150}}
\put(11701,-4861){\circle{150}}
\put(1801,-3061){\framebox(1200,2100){$A$}}
\put(3601,-3061){\framebox(1200,2100){$A^{\dagger}$}}
\put(5401,-3061){\framebox(1200,2100){$A$}}
\put(7201,-3061){\framebox(1200,2100){$A^{\dagger}$}}
\put(9001,-3061){\framebox(1200,2100){$A$}}
\put(11401,-4636){\framebox(600,750){$S$}}
\put(1201,-3961){\line( 1, 0){10200}}
\put(3301,-4036){\line( 0, 1){3000}}
\put(1201,-4111){\line( 1, 0){10200}}
\put(1201,-4261){\line( 1, 0){10200}}
\put(1201,-4411){\line( 1, 0){10200}}
\put(1201,-4561){\line( 1, 0){10200}}
\put(5131,-2881){\oval(210,210)[bl]}
\put(5131,-2191){\oval(210,210)[tl]}
\put(5071,-2881){\oval(210,210)[br]}
\put(5071,-2191){\oval(210,210)[tr]}
\put(5026,-2881){\line( 0, 1){690}}
\put(5176,-2881){\line( 0, 1){690}}
\put(8731,-2881){\oval(210,210)[bl]}
\put(8731,-2191){\oval(210,210)[tl]}
\put(8671,-2881){\oval(210,210)[br]}
\put(8671,-2191){\oval(210,210)[tr]}
\put(8626,-2881){\line( 0, 1){690}}
\put(8776,-2881){\line( 0, 1){690}}
\put(5101,-2986){\line( 0,-1){1200}}
\put(6901,-1111){\line( 0,-1){3225}}
\put(8701,-2986){\line( 0,-1){1500}}
\put(10501,-1111){\line( 0,-1){3525}}
\put(1201,-4861){\line( 1, 0){11400}}
\put(12001,-4561){\line( 1, 0){600}}
\put(12001,-4411){\line( 1, 0){600}}
\put(12001,-4261){\line( 1, 0){600}}
\put(12001,-4111){\line( 1, 0){600}}
\put(12001,-3961){\line( 1, 0){600}}
\put(10201,-2911){\line( 1, 0){2400}}
\put(10201,-2761){\line( 1, 0){2400}}
\put(10201,-2611){\line( 1, 0){2400}}
\put(10201,-2461){\line( 1, 0){2400}}
\put(10201,-2311){\line( 1, 0){2400}}
\put(10201,-2161){\line( 1, 0){2400}}
\put(10201,-1861){\line( 1, 0){2400}}
\put(10201,-1711){\line( 1, 0){2400}}
\put(10201,-1561){\line( 1, 0){2400}}
\put(10201,-2011){\line( 1, 0){2400}}
\put(10201,-1111){\line( 1, 0){2400}}
\put(10201,-1411){\line( 1, 0){2400}}
\put(11701,-4636){\line( 0,-1){300}}
\end{picture}
\end{center}
\caption{Example circuit diagram for verification procedure $B$.}
\label{fig:circuit}
\end{figure*}
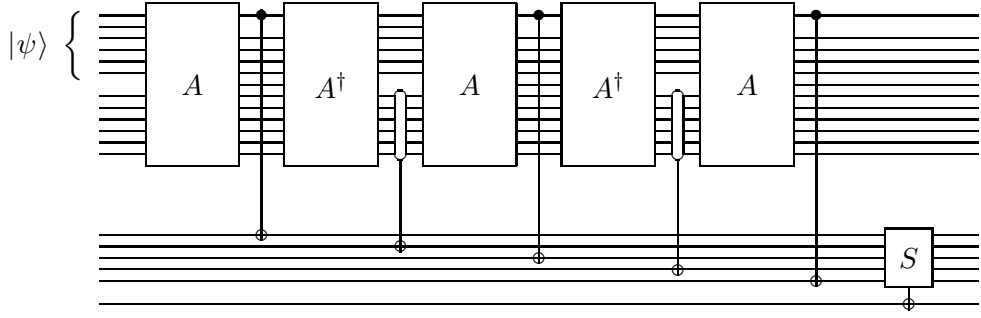
In this figure, $S$ represents the computation described in the last step of
$B$, and the last qubit rather than the first represents the output qubit to
simplify the figure.

We first consider the behavior of the verification procedure $B$
in the situation that the state $\ket{\psi}$ is an eigenvector of the
operator
\[
Q = (I_m\otimes\bra{0^k}) A^{\dagger} \Pi_1 A (I_m\otimes \ket{0^k}),
\]
with corresponding eigenvalue $p$.
We have
\[
p = \bra{\psi} Q \ket{\psi} = \left\| \Pi_1 A (\ket{\psi}\ket{0^k}) \right\|^2,
\]
and thus $p$ is the probability that the verification procedure $A$
accepts $\ket{\psi}$.
Let $\ket{\phi} = \ket{\psi}\ket{0^k}$, which implies that $\ket{\phi}$ is
an eigenvector of $\Delta_1 A^{\dagger} \Pi_1 A \Delta_1$, also having
corresponding eigenvalue $p$.
We will show that the verification procedure $B$ accepts $\ket{\psi}$
with probability
\begin{equation}\label{eq:binomial-sum}
\sum_{N\cdot \frac{a+b}{2}\leq j \leq N} \binom{N}{j} p^j (1-p)^{N-j}.
\end{equation}
Using standard Chernoff-type bounds, this probability can be shown to
be greater than $1 - 2^{-r}$ when $p \geq a$ and less than $2^{-r}$
when $p\leq b$, given the choice of $N = 8 q^2 r$.

The fact that $\ket{\psi}$ is accepted with the probability given in
equation \ref{eq:binomial-sum} will follow from the fact that the procedure
$B$ obtains each possible sequence $(z_1,\ldots,z_N)$ with probability
$p^{w(z)}(1-p)^{N - w(z)}$ for $w(z) = \sum_{i=1}^N z_i$.
This is straightforward if $p=0$ or $p=1$, so assume $0<p<1$.

Define vectors $\ket{\gamma_0}$, $\ket{\gamma_1}$, $\ket{\delta_0}$,
and $\ket{\delta_1}$ as follows:
\[
\ket{\gamma_0} = \frac{\Pi_0 A\Delta_1\ket{\phi}}{\sqrt{1-p}},\quad
\ket{\gamma_1} = \frac{\Pi_1 A\Delta_1\ket{\phi}}{\sqrt{p}},\quad
\ket{\delta_0} = \frac{\Delta_0 A^{\dagger}\Pi_1\ket{\gamma_1}}
{\sqrt{1-p}}, \quad
\ket{\delta_1} = \frac{\Delta_1 A^{\dagger}\Pi_1\ket{\gamma_1}}
{\sqrt{p}}.
\]
As $\Delta_1 A^{\dagger} \Pi_1 A \Delta_1\ket{\phi} = p\,\ket{\phi}$ and
$\ket{\phi}$ is a unit vector we have
\begin{align*}
\bra{\phi}\Delta_1 A^{\dagger} \Pi_1 A \Delta_1\ket{\phi} & = p,\\[2mm]
\bra{\phi}\Delta_1 A^{\dagger} \Pi_0 A \Delta_1\ket{\phi} & = 
\bra{\phi}\Delta_1 A^{\dagger} (I-\Pi_1) A \Delta_1\ket{\phi} = 1 - p,
\end{align*}
and thus $\ket{\gamma_0}$ and $\ket{\gamma_1}$ are unit vectors.
Moreover, as 
\[
\Pi_1 A \Delta_1 A^{\dagger} \Pi_1 \ket{\gamma_1}
= \frac{\Pi_1 A \Delta_1 \left( \Delta_1 A^{\dagger} \Pi_1 A \Delta_1\right)
\ket{\phi}}{\sqrt{p}}
= p\,\ket{\gamma_1},
\]
we have that $\ket{\delta_0}$ and $\ket{\delta_1}$ are unit vectors
by similar reasoning.
Note also that $\ket{\delta_1} = \ket{\phi}$, which follows immediately from
the fact that $\ket{\phi}$ is an eigenvector of
$\Delta_1 A^{\dagger} \Pi_1 A \Delta_1$ with eigenvalue $p$.
Based on these observations we conclude that
\begin{equation}
\label{eq:A}
\begin{split}
A\,\ket{\delta_0} & = 
-\sqrt{p}\,\ket{\gamma_0}+\sqrt{1-p}\,\ket{\gamma_1}\\[1mm]
A\,\ket{\delta_1} & = \sqrt{1-p}\,\ket{\gamma_0} + \sqrt{p}\,\ket{\gamma_1}.
\end{split}
\end{equation}
It will also be helpful to note that
\begin{equation}
\label{eq:Adagger}
\begin{split}
A^{\dagger}\,\ket{\gamma_0} & = 
-\sqrt{p}\,\ket{\delta_0}+\sqrt{1-p}\,\ket{\delta_1}\\[1mm]
A^{\dagger}\,\ket{\gamma_1} & = \sqrt{1-p}\,\ket{\delta_0} + 
\sqrt{p}\,\ket{\delta_1}
\end{split}
\end{equation}
which follows from the equations \ref{eq:A} along with the fact that $A$
is unitary.

With the above equations \ref{eq:A} and \ref{eq:Adagger} in hand, it is now
possible to calculate the probability associated with each sequence of
measurement outcomes.
The procedure $B$ begins in state $\ket{\phi} = \ket{\delta_1}$, and the 
procedure $A$ is performed.
After the measurement described by $\{\Pi_0,\Pi_1\}$ the (renormalized)
state of register $\mathsf{R}$ becomes $\ket{\gamma_0}$ or $\ket{\gamma_1}$
according to whether the outcome is 0 or 1, with associated probabilities
$1-p$ and $p$, respectively.
If instead the procedure $B$ were to start in state $\ket{\delta_0}$, the
renormalized states after measurement would be the same, but the probabilities
would be reversed; probability $p$ is associated with outcome 0 and probability
$1-p$ with outcome 1.
For the second step of the loop the situation is similar.
If the register $\mathsf{R}$ is in state $\ket{\gamma_1}$, the transformation
$A^{\dagger}$ is applied, and the state is measured with respect to the
measurement $\{\Delta_0,\Delta_1\}$, the renormalized state after measurement
will be either $\ket{\delta_1}$ or $\ket{\delta_0}$, with associated
probabilities $p$ and $1-p$.
If instead the initial state were $\ket{\gamma_0}$ rather than
$\ket{\gamma_1}$, the renormalized states after the measurement would again be
the same, but the probabilities would be reversed.
These transition probabilities are illustrated in \ref{fig:transitions}.
\begin{figure*}[t]
\vspace*{6mm}
\begin{center}
\setlength{\unitlength}{3060sp}
\begin{picture}(8158,2424)(2089,-4573)
\put(9601,-4261){\circle*{50}}
\put(9901,-4261){\circle*{50}}
\put(10201,-4261){\circle*{50}}
\put(9601,-2461){\circle*{50}}
\put(10201,-2461){\circle*{50}}
\put(9901,-2461){\circle*{50}}
\put(2101,-2761){\framebox(600,600){$\ket{\delta_1}$}}
\put(2101,-4561){\framebox(600,600){$\ket{\delta_0}$}}
\put(3901,-2761){\framebox(600,600){$\ket{\gamma_1}$}}
\put(3901,-4561){\framebox(600,600){$\ket{\gamma_0}$}}
\put(5701,-2761){\framebox(600,600){$\ket{\delta_1}$}}
\put(5701,-4561){\framebox(600,600){$\ket{\delta_0}$}}
\put(7501,-2761){\framebox(600,600){$\ket{\gamma_1}$}}
\put(7501,-4561){\framebox(600,600){$\ket{\gamma_0}$}}
\put(2701,-2461){\vector( 1, 0){1150}}
\put(2701,-4261){\vector( 1, 0){1150}}
\put(4501,-2461){\vector( 1, 0){1150}}
\put(4501,-4261){\vector( 1, 0){1150}}
\put(6301,-2461){\vector( 1, 0){1150}}
\put(6301,-4261){\vector( 1, 0){1150}}
\put(8101,-2461){\vector( 1, 0){1150}}
\put(8101,-4261){\vector( 1, 0){1150}}
\put(2701,-2461){\vector( 2,-3){1150}}
\put(2701,-4261){\vector( 2, 3){1150}}
\put(4501,-2461){\vector( 2,-3){1150}}
\put(4501,-4261){\vector( 2, 3){1150}}
\put(6301,-2461){\vector( 2,-3){1150}}
\put(6301,-4261){\vector( 2, 3){1150}}
\put(8101,-2461){\vector( 2,-3){1150}}
\put(8101,-4261){\vector( 2, 3){1150}}
\put(3301,-2350){\makebox(0,0)[lb]{$p$}}
\put(5101,-2350){\makebox(0,0)[lb]{$p$}}
\put(6826,-2350){\makebox(0,0)[lb]{$p$}}
\put(8701,-2350){\makebox(0,0)[lb]{$p$}}
\put(3226,-4520){\makebox(0,0)[lb]{$p$}}
\put(5101,-4520){\makebox(0,0)[lb]{$p$}}
\put(6826,-4520){\makebox(0,0)[lb]{$p$}}
\put(8701,-4520){\makebox(0,0)[lb]{$p$}}
\put(2441,-3150){\makebox(0,0)[lb]{$1-p$}}
\put(2441,-3750){\makebox(0,0)[lb]{$1-p$}}
\put(4241,-3150){\makebox(0,0)[lb]{$1-p$}}
\put(4241,-3750){\makebox(0,0)[lb]{$1-p$}}
\put(6041,-3150){\makebox(0,0)[lb]{$1-p$}}
\put(6041,-3750){\makebox(0,0)[lb]{$1-p$}}
\put(7841,-3150){\makebox(0,0)[lb]{$1-p$}}
\put(7841,-3750){\makebox(0,0)[lb]{$1-p$}}
\end{picture}
\end{center}
\caption{Transition probabilities for verification procedure $B$.}
\label{fig:transitions}
\end{figure*}
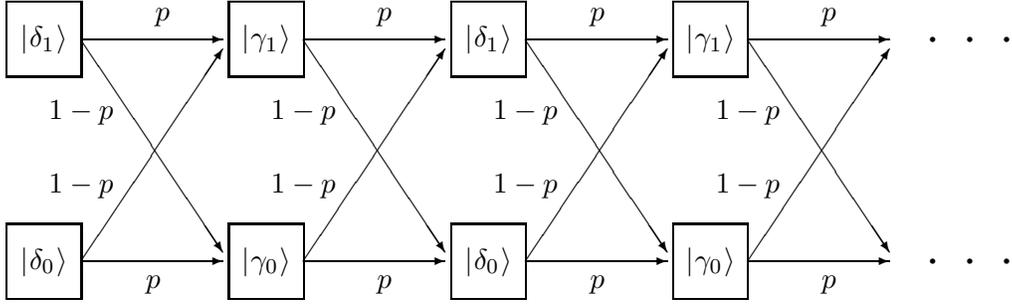
In all cases we see that the probability of obtaining the same outcome
as for the previous measurement is $p$, and the probability of the
opposite outcome is $1-p$.
The probability associated with a given sequence $z = (z_1,\ldots,z_N)$ is
therefore $p^{w(z)}(1-p)^{N-w(z)}$ as claimed, as each $z_i$ is 1 if the
measurement outcomes $y_{i-1}$ and $y_i$ are equal, and is 0 otherwise.
(Setting $y_0 = 1$ includes the first measurement outcome in this pattern.)

At this point we are ready to consider the completeness and soundness
properties of the procedure $B$.
Suppose first that the input $x$ is in $L$, which implies that the procedure
$A$ can be made to accept with probability at least~$a$.
As an arbitrary state $\ket{\psi}$ is accepted by $A$ with probability
$\bra{\psi}Q\ket{\psi}$, we therefore have 
$\bra{\psi}Q\ket{\psi} \geq a$ for some choice of $\ket{\psi}$.
Because $Q$ is positive semidefinite it is the case that
$\bra{\psi}Q\ket{\psi}$ is bounded above by the largest eigenvalue of~$Q$.
Consequently there must exist a unit eigenvector $\ket{\psi}$ of $Q$ having
associated eigenvalue $p \geq a$.
The procedure $B$ has been shown to accept such a choice of $\ket{\psi}$ with
probability at least $1-2^{-r}$ as required.

Now let us consider the soundness of the procedure $B$.
If the input $x$ is not contained in $L$, then every choice for the state
$\ket{\psi}$ causes $A$ to accept with probability at most $b$.
Therefore, every eigenvalue of the operator $Q$ is at most~$b$.
We have shown that if $\ket{\psi}$ is an eigenvector of $Q$, then the procedure
$B$ will accept $\ket{\psi}$ with probability less than $2^{-r}$.
Unfortunately, we may not assume that Merlin chooses $\ket{\psi}$ to be an
eigenvector of $Q$.
Nevertheless, the previous analysis can be extended to handle this
possibility.

Specifically, let
$\{\ket{\psi_1},\ldots,\ket{\psi_{2^m}}\}$ be a complete
orthonormal collection of eigenvectors of $Q$, with $p_j$ denoting the
eigenvalue corresponding to $\ket{\psi_j}$ for $j=1,\ldots,2^m$.
An arbitrary unit vector $\ket{\psi}$ may be written as
\[
\ket{\psi} = \sum_{j=1}^{2^m}\alpha_j\ket{\psi_j}
\]
for $\alpha_1,\ldots,\alpha_{2^m}\in\mathbb{C}$ satisfying 
$\sum_j |\alpha_j|^2 = 1$.
Given such a state $\ket{\psi}$ as input, the procedure $B$ obtains each
sequence $z = (z_1,\ldots,z_N)$ with probability
\[
\sum_{j=1}^{2^m}|\alpha_j|^2 p_j^{w(z)}(1 - p_j)^{N-w(z)}
\]
and so the probability of acceptance is
\[
\sum_{j=1}^{2^m}|\alpha_j|^2 
\sum_{N\cdot \frac{a+b}{2}\leq i \leq N} \binom{N}{i} p_j^i (1-p_j)^{N-i}
< 2^{-r}.
\]
This does not follow from linearity because measurements are nonlinear.
Instead, to see that it is indeed the case, one may repeat the analysis
given previously in somewhat more generality.
Specifically,
let $\ket{\phi_j} = \ket{\psi_j}\ket{0^k}$ and
\[
\ket{\gamma_{j,0}} = \frac{\Pi_0 A\Delta_1\ket{\phi_j}}{\sqrt{1-p_j}}, \quad
\ket{\gamma_{j,1}} = \frac{\Pi_1 A\Delta_1\ket{\phi_j}}{\sqrt{p_j}}, \quad
\ket{\delta_{j,0}} = \frac{\Delta_0 A^{\dagger}\Pi_1\ket{\gamma_{j,1}}}
{\sqrt{1-p_j}}, \quad
\ket{\delta_{j,1}} = \frac{\Delta_1 A^{\dagger}\Pi_1\ket{\gamma_{j,1}}}
{\sqrt{p_j}},
\]
for each $j = 1,\ldots,2^m$.
As before, each of these vectors is a unit vector,
$\ket{\delta_{j,1}} = \ket{\phi_j}$, and
\begin{align*}
A\,\ket{\delta_{j,0}} & = 
-\sqrt{p_j}\,\ket{\gamma_{j,0}}+\sqrt{1-p_j}\,\ket{\gamma_{j,1}},\\[1mm]
A\,\ket{\delta_{j,1}} & = \sqrt{1-p_j}\,\ket{\gamma_{j,0}} + \sqrt{p_j}\,
\ket{\gamma_{j,1}},\\[1mm]
A^{\dagger}\,\ket{\gamma_{j,0}} & = 
-\sqrt{p_j}\,\ket{\delta_{j,0}}+\sqrt{1-p_j}\,\ket{\delta_{j,1}},\\[1mm]
A^{\dagger}\,\ket{\gamma_{j,1}} & = 
\sqrt{1-p_j}\,\ket{\delta_{j,0}} + \sqrt{p_j}\,\ket{\delta_{j,1}}.
\end{align*}
Moreover, each of the sets $\{\ket{\gamma_{j,0}}\}$, $\{\ket{\gamma_{j,1}}\}$,
$\{\ket{\delta_{j,0}}\}$, and $\{\ket{\delta_{j,1}}\}$ is an orthonormal set.
Because of this fact, when $B$ is performed on the state $\ket{\psi}$,
a similar pattern to the single eigenvector case arises independently for each
eigenvector $\ket{\psi_j}$.
This results in the stated probability of acceptance, which completes the
proof.
\end{proof}


\subsection*{Applications of strong error reduction}
\label{sec:applications}

Two applications of \ref{theorem:amplify} will now be discussed.
The first is a simplified proof that $\mathrm{QMA}$ is contained in the class
$\mathrm{PP}$.

\begin{theorem}
\label{theorem:QMA_in_PP}
$\mathrm{QMA}\subseteq\mathrm{PP}$.
\end{theorem}
\begin{proof}
Let $L\subseteq\Sigma^{\ast}$ be a language in $\mathrm{QMA}$.
By \ref{theorem:amplify} there exists a function $m\in\mathit{poly}$
such that
\[
L\in\mathrm{QMA}_m\left(1-2^{-(m+2)},2^{-(m+2)}\right).
\]
Let $A$ be a verification procedure that witnesses this fact.
Specifically, each circuit $A_x$ acts on $k + m$ qubits,
for some $k\in\mathit{poly}$, and satisfies the following.
If $x\in L$, then there exists an $m$ qubit state $\ket{\psi}$ such that
\[
\op{Pr}[\mbox{$A_x$ accepts $\ket{\psi}$}] \geq 1 - 2^{-m-2},
\]
while if $x\not\in L$, then
\[
\op{Pr}[\mbox{$A_x$ accepts $\ket{\psi}$}] \leq 2^{-m-2}
\]
for every $m$ qubit state $\ket{\psi}$.

For each $x\in\Sigma^{\ast}$, define a $2^m\times 2^m$ matrix $Q_x$ as
\[
Q_x = \left( I_m \otimes \bra{0^k}\right) A_x^{\dagger} \Pi_1
A_x \left(I_m \otimes \ket{0^k}\right).
\]
Each $Q_x$ is positive semidefinite, and
$\bra{\psi} Q_x \ket{\psi} = \op{Pr}[\mbox{$A_x$ accepts $\ket{\psi}$}]$
for any unit vector $\ket{\psi}$ on $m$ qubits.
The maximum probability with which $A_x$ can be made to accept is
the largest eigenvalue of $Q_x$.
Because the trace of a matrix is equal to the sum of its eigenvalues
and all eigenvalues of $Q_x$ are nonnegative,
it follows that if $x\in L$, then
$\op{tr}(Q_x) \geq 1 - 2^{-m-2}\geq 3/4$,
while if $x\not\in L$, then
$\op{tr}(Q_x) \leq 2^{m} 2^{-m-2}\leq 1/4$.

Now, based on a straightforward modification of the method of \cite{FortnowR99}
discussed previously, we have that there exists a polynomially-bounded
$\mathrm{FP}$ function $g$ and $\mathrm{GapP}$ functions $f_1$ and $f_2$ such
that the real and imaginary parts of the entries of $Q_x$ are represented by
$f_1$, $f_2$, and $g$ in the sense that
\[
\Re(Q_x[i,j]) = \frac{f_1(x,i,j)}{2^{g(x)}}\;\;\;\;\text{and}\;\;\;\;
\Im(Q_x[i,j]) = \frac{f_2(x,i,j)}{2^{g(x)}}
\]
for $0\leq i,j<2^{m}$.
Define
\[
h(x) = \sum_{i = 0}^{2^{m}-1} f_1(x,i,i).
\]
Because $\mathrm{GapP}$ functions are closed under exponential sums,
we have $h\in\mathrm{GapP}$.
It holds that $h(x) = 2^{g(x)} \op{tr}(Q_x)$, and therefore
\[
x\in L \;\Rightarrow\; h(x)\geq \frac{3}{4}\,2^{g(x)}\quad\text{and}\quad
x\not\in L \;\Rightarrow\; h(x)\leq \frac{1}{4}\,2^{g(x)}.
\]
Because $2^{g(x)}$ is an FP function, it follows that $2h(x) - 2^{g(x)}$ is
a GapP function that is positive if $x\in L$ and negative if $x\not\in L$.
Thus, $L\in\mathrm{PP}$ as required.
\end{proof}

\begin{remark}
A simple modification of the above proof yields
$\mathrm{QMA}\subseteq\mathrm{A}_0\mathrm{PP}$.
Specifically, the $\mathrm{GapP}$ function $2 h$ and the $\mathrm{FP}$
function $2^{g(x)}$ satisfy the required properties to prove
$L\in\mathrm{A}_0\mathrm{PP}$, namely
\[
x\in L \;\Rightarrow\; 2 h(x)\geq 2^{g(x)}\quad\text{and}\quad
x\not\in L \;\Rightarrow\; 2 h(x)\leq \frac{1}{2}\,2^{g(x)}.
\]
\end{remark}

The second application concerns one-message quantum Arthur-Merlin games where
Merlin sends only a logarithmic number of qubits to Arthur.
Classical one-message Arthur-Merlin games with logarithmic-length messages
from Merlin to Arthur are obviously equivalent in power to $\mathrm{BPP}$,
because Arthur could simply search through all possible messages in polynomial
time in lieu of interacting with Merlin.
In the quantum case, however, this argument does not work, as
one may construct exponentially large sets of pairwise nearly-orthogonal
quantum states on a logarithmic number of qubits, such as those used
in quantum fingerprinting \cite{BuhrmanC+01}.
Nevertheless, logarithmic length quantum messages can be shown to be
useless in the context of $\mathrm{QMA}$ using a different method, based
on the strong error reduction property of $\mathrm{QMA}$ proved above.

For $a,b:\mathbb{N}\rightarrow[0,1]$ define
$\mathrm{QMA}_{\mathrm{log}}(a,b)$ to be the class of all languages contained
in $\mathrm{QMA}_m(a,b)$ for $m(n) = O(\log n)$, and let
\[
\mathrm{QMA}_{\mathrm{log}} = \mathrm{QMA}_{\mathrm{log}}(2/3,1/3).
\]
The choice of the constants 2/3 and 1/3 is arbitrary, which follows from
\ref{theorem:amplify}.

\begin{theorem}
$\mathrm{QMA}_{\mathrm{log}} = \mathrm{BQP}$.
\end{theorem}

\begin{proof}
The containment 
$\mathrm{BQP}\subseteq\mathrm{QMA}_{\mathrm{log}}$ is trivial, so it
suffices to prove $\mathrm{QMA}_{\mathrm{log}} \subseteq \mathrm{BQP}$.
Assume $L\in\mathrm{QMA}_m$ for $m$ logarithmic, and assume $A$ is a
$\mathrm{QMA}$ verification procedure that witnesses this fact and has
completeness and soundness error less than $2^{-(m+2)}$.
Let
\[
Q_x = \left(I_m \otimes \bra{0^k}\right)A_x^{\dagger}\Pi_1 A_x
\left(I_m \otimes \ket{0^k}\right).
\]
Similar to the proof of \ref{theorem:QMA_in_PP}, we have
\[
x\in L\;\Rightarrow\;\op{tr}(Q_x)\geq 3/4,\;\;\;\;\;
x\not\in L\;\Rightarrow\;\op{tr}(Q_x)\leq 1/4.
\]

We will describe a polynomial-time quantum algorithm $B$ that
decides $L$ with bounded error.
The algorithm $B$ simply constructs a totally mixed state over $m$ qubits
and runs the verification procedure $A$ using this state in place of Merlin's
message.
Running the verification procedure on the totally mixed state is equivalent
to running the verification procedure on $m$ qubits initialized to some
uniformly generated standard basis state, which is straightforward to
simulate using Hadamard transforms and reversible computation.
The totally mixed state on $m$ qubits corresponds to the density
matrix $2^{-m}I_m$, from which it follows that the probability of
acceptance of $B$ is given by
\[
\op{Pr}[\mbox{$B$ accepts $x$}] = \op{tr}\left(Q_x\, 2^{-m} I_m\right)
= 2^{-m}\op{tr}(Q_x).
\]
Given that $m$ is logarithmic in $|x|$, we have that the probabilities
with which $B$ accepts inputs $x\in L$ and inputs $x\not\in L$ are
bounded away from one another by the reciprocal of some polynomial.
This difference can be amplified by standard methods, implying that
$L \in \mathrm{BQP}$.
\end{proof}


\section{QAM}
\label{sec:QAM}


A $\mathrm{QAM}$ verification procedure $A$ consists of a polynomial-time
generated family
\[
\left\{A_{x,y}\,:\,x\in\Sigma^{\ast},\,y\in\Sigma^{s(|x|)}\right\}
\]
of quantum circuits together with functions $m,s\in\mathit{poly}$.
As for $\mathrm{QMA}$ verification procedures, each circuit
$A_{x,y}$ acts on two collections of qubits: $m(|x|)$ qubits sent by Merlin and
$k(|x|)$ qubits corresponding to Arthur's workspace.
The notion of a circuit $A_{x,y}$ accepting a message $\ket{\psi}$
is defined in the same way as for $\mathrm{QMA}$.
In the present case, the string $y$ corresponds to a sequence of coin-flips
sent by Arthur to Merlin, on which Merlin's message may depend.

\begin{definition}
The class $\mathrm{QAM}(a,b)$ consists of all languages
$L\subseteq\Sigma^{\ast}$ for which there exists a $\mathrm{QAM}$ verification
procedure $A$ satisfying the following conditions.
\begin{mylist}{\parindent}
\item[1.]
If $x\in L$ then there exists a collection of states $\{\ket{\psi_y}\}$
on $m$ qubits such that
\[
\frac{1}{2^s}\sum_{y\in\Sigma^{s}}
\op{Pr}[\mbox{$A_{x,y}$ accepts $\ket{\psi_y}$}]
\:\geq\:a.
\]

\item[2.]
If $x\not\in L$ then for every collection of states $\{\ket{\psi_y}\}$ on
$m$ qubits it holds that
\[
\frac{1}{2^s}\sum_{y\in\Sigma^{s}}
\op{Pr}[\mbox{$A_{x,y}$ accepts $\ket{\psi_y}$}]
\:\leq\:b.
\]
\end{mylist}
Similar to $\mathrm{QMA}$, one may consider the cases where $a$ and $b$ are
constants or functions of $n = |x|$, and in the case that $a$ and $b$ are
functions of the input length it is assumed that $a(n)$ and $b(n)$ can be
computed deterministically in time polynomial in $n$.
Also as before, let $\mathrm{QAM} = \mathrm{QAM}(2/3,1/3)$.
\end{definition}


\subsection*{Error reduction for QAM}

The first fact about $\mathrm{QAM}$ that we prove is that completeness and
soundness errors may be reduced by running many copies of a given game
in parallel.
The proof is similar in principle to the proof of Lemma 14.1 in
\cite{KitaevS+02}, which corresponds to our \ref{theorem:Kitaev1}.

\begin{theorem}
\label{theorem:QAM_error_reduction}
Let $a,b:\mathbb{N}\rightarrow[0,1]$ and $q\in\mathit{poly}$ satisfy
\[
a(n) - b(n)\geq \frac{1}{q(n)}
\]
for all $n\geq \mathbb{N}$.
Then 
$\mathrm{QAM}(a,b) \subseteq \mathrm{QAM}(1-2^{-r},2^{-r})$
for every $r\in\mathit{poly}$.
\end{theorem}

\begin{proof}
Let $L\in\mathrm{QAM}(a,b)$, and let $A$ be a $\mathrm{QAM}$ verification
procedure witnessing this fact.
We consider a new $\mathrm{QAM}$ verification procedure that corresponds to
playing the game described by $\{A_{x,y}\}$ in parallel $N$ times.
The new procedure accepts if and only if the number of acceptances of the
original game is at least $N\cdot\frac{a+b}{2}$.
Although Merlin is not required to play the repetitions independently,
we will show that playing the repetitions independently in fact gives
him an optimal strategy.
The theorem then follows by choosing an appropriately large value of
$N$ and applying a Chernoff-type bound.

Assume hereafter that the input $x$ is fixed, and define
\begin{align*}
Q_y^{(0)} & = (I\otimes\bra{0^k})A_{x,y}^{\dagger}\Pi_0 A_{x,y}
(I\otimes\ket{0^k}),\\
Q_y^{(1)} & = (I\otimes\bra{0^k})A_{x,y}^{\dagger}\Pi_1 A_{x,y}
(I\otimes\ket{0^k})
\end{align*}
for each $y\in\Sigma^s$.
We have $Q_y^{(1)} = I - Q_y^{(0)}$, and consequently $Q_y^{(0)}$ and
$Q_y^{(1)}$ share a complete set of orthonormal eigenvectors.
Let $\{\ket{\psi_{y,1}},\ldots,\ket{\psi_{y,2^m}}\}$ be such a set, and let
\[
p^{(z)}_{y,1},\ldots,p^{(z)}_{y,2^m}
\]
be the corresponding eigenvalues
for $Q_y^{(z)}$, $z\in\{0,1\}$.
As $Q_y^{(0)}$ and $Q_y^{(1)}$ are positive semidefinite and sum to the
identity, $p^{(0)}_{y,i}$ and $p^{(1)}_{y,i}$ are nonnegative real numbers with
$p^{(0)}_{y,i} + p^{(1)}_{y,i} = 1$ for each $y$ and $i$.
Assume without loss of generality that the eigenvectors and eigenvalues
are ordered in such a way that
\[
p^{(1)}_{y,1}\geq \cdots \geq p^{(1)}_{y,2^m}.
\]
This implies that the maximum acceptance probability of $A_{x,y}$
is $p^{(1)}_{y,1}$.

Under the assumption that Arthur's coin-flips for the $N$ repetitions
are given by strings $y_1,\ldots,y_N\in\Sigma^s$, if Merlin plays the
repetitions independently, and optimally for each repetition, his probability
of convincing Arthur to accept is
\begin{equation}
\label{eq:optimal}
\sum_{\stackrel{\scriptstyle{z_1,\ldots,z_N\in\Sigma}}
{\scriptstyle{z_1+\cdots +z_N\geq N\cdot\frac{a+b}{2}}}}
p_{y_1,1}^{(z_1)}\cdots p_{y_N,1}^{(z_N)}.
\end{equation}
Without any assumption on Merlin's strategy, the maximum probability with
which Merlin can win $N\cdot\frac{a+b}{2}$ repetitions of the original game
when Arthur's coin-flips are given by $y_1,\ldots,y_N$ is equal to
the largest eigenvalue of
\begin{equation}
\label{eq:optimal2}
\sum_{\stackrel{\scriptstyle{z_1,\ldots,z_N\in\Sigma}}
{\scriptstyle{z_1+\cdots +z_N\geq N\cdot\frac{a+b}{2}}}}
Q_{y_1}^{(z_1)}\otimes \cdots \otimes Q_{y_N}^{(z_N)}.
\end{equation}
Therefore, to prove the proposition it suffices to show that these
quantities are equal.

All of the summands in equation \ref{eq:optimal2}
share the complete set of orthonormal eigenvalues given by
\[
\left\{
\ket{\psi_{y_1,i_1}}\cdots\ket{\psi_{y_N,i_N}}\,:\,i_1,\ldots,i_N\in
\{1,\ldots,2^m\}\right\},
\]
and so this set also describes a complete set of orthonormal eigenvectors
of the sum.
The eigenvalue associated with
$\ket{\psi_{y_1,i_1}}\cdots\ket{\psi_{y_N,i_N}}$ is
\begin{equation}
\label{eq:qam_amplify1}
\sum_{\stackrel{\scriptstyle{z_1,\ldots,z_N\in\Sigma}}
{\scriptstyle{z_1+\cdots +z_N\geq N\cdot\frac{a+b}{2}}}}
p_{y_1,i_1}^{(z_1)}\cdots p_{y_N,i_N}^{(z_N)}.
\end{equation}

Define $u_1(X) = X$, $u_0(X) = 1 - X$, and let
\[
f(X_1,\ldots,X_N)=\!\!
\sum_{\stackrel{\scriptstyle{z_1,\ldots,z_N\in\Sigma}}
{\scriptstyle{z_1+\cdots +z_N\geq N\cdot\frac{a+b}{2}}}}\!\!
u_{z_1}(X_1)\cdots u_{z_N}(X_N).
\]
The quantity in equation \ref{eq:qam_amplify1} is equal to
\[
f\left(p_{y_1,i_1}^{(1)},\ldots,p_{y_N,i_N}^{(1)}\right).
\]
The function $f$ is multi-linear and nondecreasing in each variable everywhere
on the unit hypercube.
Thus, the maximum of the quantity in equation \ref{eq:qam_amplify1} is
\[
f\left(p_{y_1,1}^{(1)},\ldots,p_{y_N,1}^{(1)}\right),
\]
which is equal to the quantity in equation \ref{eq:optimal}.
This completes the proof.
\end{proof}


\subsection*{An upper bound on QAM}

We now observe that the upper bound 
\[
\mathrm{QAM}\subseteq\mathrm{BP}\cdot\mathrm{PP}
\]
holds.
The following fact concerning the maximum probabilities of acceptance
of $A_{x,y}$ for random $y$ will be used.
Here we let $\mu(A_{x,y})$ denote the maximum probability that
$A_{x,y}$ can be made to accept (maximized over all choices of Merlin's message
$\ket{\psi_y}$).

\begin{prop}
\label{prop:markov}
Suppose that
\[
\left\{A_{x,y}\,:\,x\in\Sigma^{\ast},\,y\in\Sigma^{s(|x|)}\right\}
\]
is a $\mathrm{QAM}$ verification procedure for a language $L$ that has
completeness and soundness errors bounded by 1/9.
Then for any $x\in\Sigma^{\ast}$ and for $y\in\Sigma^{s}$ chosen uniformly at
random,
\begin{align*}
x\in L & \Rightarrow \op{Pr}[\mu(A_{x,y})\geq 2/3]\:\geq\:
2/3\\[2mm]
x\not\in L & \Rightarrow \op{Pr}[\mu(A_{x,y})\leq 1/3] 
\:\geq\: 2/3.
\end{align*}
\end{prop}
\begin{proof}
Suppose that $x\in L$.
Let $z(y) =  1 - \mu(A_{x,y})$,
and let $Z$ be a random variable whose value is $z(y)$ for a uniformly
chosen $y\in\Sigma^{s}$.
The assumption of the proposition implies that $E[Z] \leq 1/9$.
By Markov's inequality we have
\[
\op{Pr}[Z > 1/3] \leq \frac{E[Z]}{1/3} \leq 1/3,
\]
and therefore
\[
\op{Pr}[\mu(A_{x,y})\geq 2/3] = \op{Pr}[Z \leq 1/3] \geq 2/3.
\]
The proof for $x\not\in L$ is similar.
\end{proof}

\begin{theorem}
\label{theorem:QAM_in_BP.PP}
$\mathrm{QAM}\subseteq\mathrm{BP}\cdot\mathrm{PP}$.
\end{theorem}

\begin{proof}
Let $L \in \mathrm{QAM}$, and let
\[
A = \left\{A_{x,y}\,:\,x\in\Sigma^{\ast},\,y\in\Sigma^{s(|x|)}\right\}
\]
be a $\mathrm{QAM}$ verification procedure for $L$ with completeness and
soundness errors bounded by 1/9.
Such a procedure exists by \ref{theorem:QAM_error_reduction}.
By a straightforward modification of the proof of \ref{theorem:QMA_in_PP},
one may conclude that there exists a language $K\in\mathrm{PP}$ such that
\begin{align*}
\mu(A_{x,y})\geq 2/3 & \Rightarrow (x,y) \in K,\\[2mm]
\mu(A_{x,y})\leq 1/3 & \Rightarrow (x,y) \not\in K.
\end{align*}
It is possible that $\mu(A_{x,y})\in (1/3,2/3)$ for some values
of $y$, but in this case no requirement is made on whether or not $(x,y)\in K$.
The theorem now follows from \ref{prop:markov}.
\end{proof}


\section{QMAM}
\label{sec:QMAM}

A $\mathrm{QMAM}$ verification procedure $A$ consists of
a polynomial-time generated family
\[
\left\{A_{x,y}\,:\,x\in\Sigma^{\ast},\,y\in\Sigma^{s(|x|)}\right\}
\]
of quantum circuits, together with functions $m_1,m_2,s\in\mathit{poly}$.
The functions $m_1$ and $m_2$ specify the number of qubits in Merlin's
first and second messages to Arthur, while $s$ specifies the number of
random bits Arthur sends to Merlin.
Each circuit $A_{x,y}$ acts on $m_1(|x|) + m_2(|x|) + k(|x|)$ qubits, where
as before $k(|x|)$ denotes the number of qubits corresponding to Arthur's
workspace.

In the $\mathrm{QMAM}$ case, it becomes necessary to discuss possible actions
that Merlin may perform rather than just discussing states that he may send.
This is because Merlin's strategy could involve preparing some quantum state,
sending part of that state to Arthur on the first message, and transforming
the part of that state he did not send to Arthur (after receiving Arthur's
coin-flips) in order to produce his second message.

\begin{definition}
A language $L\subseteq\Sigma^{\ast}$ is in $\mathrm{QMAM}(a,b)$ if there exists
a $\mathrm{QMAM}$ verification procedure $A$ such that the following conditions
are satisfied.
\begin{mylist}{\parindent}
\item[1.]
If $x\in L$ then for some $l$ there exists a quantum state
$\ket{\psi}$ on $m_1 + m_2 + l$ qubits and a collection
of unitary operators $\{U_y\,:\,y\in\Sigma^{s}\}$ acting on $m_2 + l$ qubits
such that
\[
\frac{1}{2^s}\sum_{y\in\Sigma^{s}}
\op{Pr}[\mbox{$A_{x,y}$ accepts $(I_{m_1}\otimes U_y)\ket{\psi}$}]\:\geq\:a.
\]

\item[2.]
If $x\not\in L$ then for every $l$, every quantum state $\ket{\psi}$ on
$m_1 + m_2 + l$ qubits, and every collection of unitary operators
$\{U_y\,:\,y\in\Sigma^{s}\}$ acting on $m_2 + l$ qubits,
\[
\frac{1}{2^s}\sum_{y\in\Sigma^{s}}
\op{Pr}[\mbox{$A_{x,y}$ accepts $(I_{m_1}\otimes U_y)\ket{\psi}$}]
\:\leq\:b.
\]
\end{mylist}
The same assumptions regarding $a$ and $b$ apply in this case as in
the $\mathrm{QMA}$ and $\mathrm{QAM}$ cases.
\end{definition}

\noindent
In the above definition, the circuit $A_{x,y}$ is acting on $m_1+m_2$ qubits
sent by Merlin in addition to Arthur's $k$ workspace qubits, while
$(I_{m_1}\otimes U_y)\ket{\psi}$ is a state on $m_1 + m_2 + l$ qubits.
It is to be understood that the last $l$ qubits of
$(I_{m_1}\otimes U_y)\ket{\psi}$ remain in Merlin's possession, so $A_{x,y}$
is effectively tensored with the identity acting on these qubits.


\subsection*{Equivalence of QMAM and QIP}
\label{sec:QMAM=QIP}

We now  prove $\mathrm{QMAM} = \mathrm{QIP}$.
Because quantum Arthur-Merlin games are a restricted form of quantum
interactive proof systems, $\mathrm{QMAM}\subseteq\mathrm{QIP}$ is obvious.
To prove the opposite containment, we will require the following lemmas.
The first lemma is a corollary of Uhlmann's Theorem (see \cite{NielsenC00}).
\begin{lemma}
\label{lemma:cor_of_Uhlmann}
Suppose the pair of registers $(\mathsf{V},\mathsf{M})$ is in a mixed
state for which the reduced state of $\,\mathsf{V}$ is~$\sigma$.
If the pair $(\mathsf{V},\mathsf{M})$ is measured with respect to a binary
valued measurement described by orthogonal projections
$\{\Lambda_0,\Lambda_1\}$,
then the probability of obtaining the outcome 1 is at most
$F(\sigma,\rho)^2$ for some $\rho\in\mathcal{S}_{\mathsf{V}}(\Lambda_1)$.
\end{lemma}
\noindent
The second lemma is a simple property of the fidelity function.
\begin{lemma}[\cite{NayakS02,SpekkensR02}]
\label{lemma:fidelity_sum}
For any choice of density matrices $\rho$, $\xi$, and $\sigma$, we have
\[
F(\rho,\sigma)^2 + F(\sigma,\xi)^2 \leq 1 + F(\rho,\xi).
\]
\end{lemma}

\begin{theorem}
Let $L\in\mathrm{QIP}$ and let $r\in\mathit{poly}$.
Then $L$ has a three message quantum Arthur-Merlin game with completeness
error 0 and soundness error at most $1/2 + 2^{-r}$.
Moreover, in this quantum Arthur-Merlin game, Arthur's message consists of a
single coin-flip.
\end{theorem}

\begin{proof}
Let $L\in\mathrm{QIP}$, which implies that $L$ has a three-message quantum
interactive proof system with completeness error 0 and soundness error
$\varepsilon(n) = 2^{-2 r(n)}$ on inputs of length $n$.

Consider a $\mathrm{QMAM}$ verification procedure $A$ that corresponds to the
following actions for Arthur.
(It will be assumed that the input $x$ is fixed, and it will be clear that
the family of quantum circuits corresponding to this verification procedure
can be generated in polynomial-time given that the same is true of
the verifier being simulated.)
\begin{mylist}{\parindent}
\item[1.] Receive register $\mathsf{V}$ from Merlin.

\item[2.] Flip a fair coin and send the result to Merlin.

\item[3.] Receive register $\mathsf{M}$ from Merlin.
If the coin flipped in step 2 was {\sc heads}, apply $V_2$ to
$(\mathsf{V},\mathsf{M})$ and accept if the first qubit of $\mathsf{V}$
(i.e., the output qubit of the quantum interactive proof system) is 1,
otherwise reject.
If the coin in step 2 was {\sc tails}, apply $V_1^{\dagger}$ to
$(\mathsf{V},\mathsf{M})$ and accept if all qubits of $\mathsf{V}$ are
set to 0, otherwise reject.
\end{mylist}

Suppose first that $x\in L$, so that some prover, whose actions are described
by a state $\ket{\psi}$ and a unitary operator $U$ can convince $V$ to accept
with certainty.
Then Merlin can convince Arthur to accept with certainty as follows:
\begin{mylist}{\parindent}
\item[1.]
Prepare state $\ket{0^k}$ in register $\mathsf{V}$ and state
$\ket{\psi}$ in registers $(\mathsf{M},\mathsf{P})$.
Apply $V_1$ to registers $(\mathsf{V},\mathsf{M})$, and send $\mathsf{V}$ to
Arthur.
\item[2.]
If Arthur flips {\sc heads}, apply $U$ to $(\mathsf{M},\mathsf{P})$ and
send $\mathsf{M}$ to Arthur.
If Arthur flips {\sc tails}, send $\mathsf{M}$ to Arthur without
applying $U$.
\end{mylist}

Now assume $x\not\in L$, so that no prover can convince $V$ to accept
with probability exceeding $\varepsilon$.
Suppose that the reduced density matrix of register $\mathsf{V}$ sent
by Merlin is $\sigma$.
By \ref{lemma:cor_of_Uhlmann} and \ref{lemma:fidelity_sum},
the probability that Arthur can be made to accept is at most
\[
\frac{1}{2}F(\rho,\sigma)^2 + \frac{1}{2}F(\xi,\sigma)^2
\leq \frac{1}{2} + \frac{1}{2} F(\rho,\xi)
\]
maximized over
$\rho\in\mathcal{S}_{\mathsf{V}}(V_1 \Delta_1 V_1^{\dagger})$
and $\xi\in\mathcal{S}_{\mathsf{V}}(V_2^{\dagger} \Pi_1 V_2)$.
By \ref{prop:max-accept} this probability is at most
\[
\frac{1}{2} + \frac{\sqrt{\varepsilon}}{2}\leq \frac{1}{2} + 2^{-r(|x|)},
\]
which completes the proof.
\end{proof}

\begin{cor}
For any function $r\in\mathit{poly}$
we have $\mathrm{QIP}\subseteq\mathrm{QMAM}(1,1/2 + 2^{-r})$.
\end{cor}

\subsection*{Error reduction for QMAM}

Now, suppose that we have a $\mathrm{QMAM}$ protocol for a language $L$ with
perfect completeness and soundness error $b$, and we repeat the protocol
$N$ times in parallel, accepting if and only if all $N$ of the repetitions
accept.
It is clear that this resulting protocol has perfect completeness,
because Merlin can play optimally for each parallel repetition independently
and achieve an acceptance probability of 1 for any $x\in L$.
In the case that $x\not\in L$, Merlin can gain no advantage whatsoever
over playing the repetitions independently, and so the soundness error
decreases to $b^N$ as we would hope.
This follows from the fact that the same holds for arbitrary three-message
quantum interactive proof systems \cite{KitaevW00}, of which
three-message quantum Arthur-Merlin games are a restricted type.
This implies the following corollary.

\begin{cor}
For any function $r\in\mathit{poly}$ we have
$\mathrm{QIP} = \mathrm{QMAM}(1,2^{-r})$.
\end{cor}


\subsection*{More than three messages}

Finally, we note that one may define quantum Arthur-Merlin games having any
polynomial number of messages in a similar way to three-message quantum
Arthur-Merlin games.
Such games are easily seen to be equivalent in power to three-message
quantum Arthur-Merlin games.
Specifically, polynomial-message quantum Arthur-Merlin games will
be special cases of quantum interactive proof systems, and can therefore be
parallelized to three-message interactive proofs and simulated by
three-message quantum Arthur-Merlin games as previously described.


\section{Open questions}
\label{sec:conclusion}

Many interesting questions about quantum Arthur-Merlin games remain
unanswered, including the following questions.

\begin{mylist}{\parindent}
\item[$\bullet$]
Are there interesting examples of problems in $\mathrm{QMA}$ or
$\mathrm{QAM}$ that are not known to be in $\mathrm{AM}$?
A similar question may be asked for $\mathrm{QMAM}$ vs.~$\mathrm{PSPACE}$.

\item[$\bullet$]
The question of whether there exists an oracle relative to which
$\mathrm{BQP}$ is outside of the polynomial-time hierarchy appears to be a
difficult problem.
In fact it is currently not even known if there is an oracle relative to
which $\mathrm{BQP}\not\subseteq \mathrm{AM}$.
Is there an oracle relative to which $\mathrm{QMA}$ or $\mathrm{QAM}$ is not
contained in $\mathrm{AM}$?
If so, what about $\mathrm{QMA}$ or $\mathrm{QAM}$ versus $\mathrm{PH}$?
Such results might shed some light on the problem of 
$\mathrm{BQP}$ versus the polynomial-time hierarchy.

\item[$\bullet$]
\cite{NisanW94} proved $\mathrm{almost}$-$\mathrm{NP} = \mathrm{AM}$.
Is it the case that $\mathrm{almost}$-$\mathrm{QMA} = \mathrm{QAM}$?

\end{mylist}


\subsection*{Acknowledgements}
Thanks to Dorit Aharonov, Oded Regev, and Umesh Vazirani for their comments
on error reduction for $\mathrm{QMA}$, Ashwin Nayak for helpful references, and
Alexei Kitaev for discussions about quantum proof systems.
This research was supported by Canada's NSERC,
the Canadian Institute for Advanced Research (CIAR), and the
Canada Research Chairs program.


\bibliographystyle{alpha}

\end{document}